# Algorithms for Routing an Unmanned Aerial Vehicle in the presence of Refueling Depots

Kaarthik Sundar[1], Sivakumar Rathinam[2]

*Abstract*—We consider a single Unmanned Aerial Vehicle (UAV) routing problem where there are multiple depots and the vehicle is allowed to refuel at any depot. The objective of the problem is to find a path for the UAV such that each target is visited at least once by the vehicle, the fuel constraint is never violated along the path for the UAV, and the total fuel required by the UAV is a minimum. We develop an approximation algorithm for the problem, and propose fast construction and improvement heuristics to solve the same. Computational results show that solutions whose costs are on an average within 1.4% of the optimum can be obtained relatively fast for the problem involving 5 depots and 25 targets.

**Note to Practitioners**

The motivation for this paper stems from the need to develop path planning algorithms for small UAVs with resource constraints. Small autonomous UAVs are seen as ideal platforms for many monitoring applications. Small UAVs can fly at low altitudes and can avoid obstacles or threats at low altitudes more easily. These vehicles can also be hand launched by an individual without any reliance on a specific type of terrain. Even though there are several advantages with using small platforms, they also come with other resource constraints due to their size and limited payload. This article addresses a path planning problem involving a small UAV with fuel constraints, and presents fast and efficient algorithms for finding good feasible solutions.

## I. INTRODUCTION

Path planning for small Unmanned Aerial Vehicles (UAVs) is one of the research areas that has received significant attention in the last decade. Small UAVs have already been field tested in civilian applications such as wild-fire management [1], weather and hurricane monitoring [2], [3], and pollutant estimation [4] where the vehicles are used to collect relevant sensor information and transmit the information to the ground (control) stations for further processing. Compared to large UAVs, small UAVs are relatively easier to operate and are significantly cheaper. Small UAVs can fly at low altitudes and can avoid obstacles or threats at low altitudes more easily. Even in military applications, small vehicles [5] are used frequently for intelligence gathering and damage assessment as they are easier to fly and can be hand launched by an individual without any reliance on a runway or a specific type of terrain.

Even though there are several advantages with using small platforms, they also come with other resource constraints due to their size and limited payload. As small UAVs typically have fuel constraints, it may not be possible for an UAV to complete a surveillance mission before refueling at one of the depots. For example, consider a typical surveillance mission where a vehicle starts at a depot and is required to visit a set of targets. To complete this mission, the vehicle may have to start at the depot, visit a subset of targets and then reach one of the depots for refueling before starting a new path. One can reasonably assume that once the UAV reaches a depot, it will be refueled to full capacity before it leaves again for visiting any remaining targets. If the goal is to visit each of the given targets at least once, then the UAV may have to repeatedly visit some depots in order to refuel again before visiting all the targets. In this scenario, the following **Fuel Constrained, UAV Routing Problem** (FCURP) naturally arises: Given a set of targets, depots, and an UAV where the vehicle is initially stationed at one of the depots, find a path for the UAV such that each target is visited at least once by the vehicle, the fuel constraint is never violated along the path for the UAV, and the travel cost for the vehicle is a minimum. The travel cost is defined as the total fuel consumed by the vehicle as it traverses its path. Please refer to figure 1 for an illustration of this problem. If the UAV is modeled as a Dubins' vehicle [6] with a bound on its turning radius, it is possible that the travel costs are asymmetric. Asymmetry means that the cost of traveling from target $A$ to target $B$ may not be equal to the cost of traveling from target $B$ to target $A$.

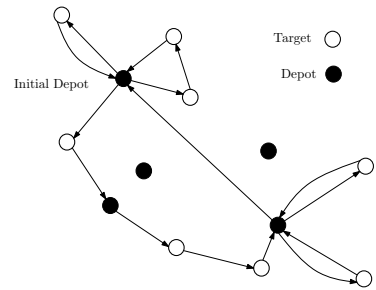

Fig. 1. A possible path for the UAV which visits all the targets while visiting some depots for refueling. Note that a depot can be visited any number of times for refueling while some depots may not be visited at all.

FCURP is a generalization of the Asymmetric Traveling Salesman Problem (ATSP) and is NP-Hard. Therefore, the main objective of this article is to develop an approximation algorithm and heuristics to solve the FCURP. A $\alpha$-approximation algorithm for an optimization problem is an algorithm that runs in polynomial time and finds a feasible solution whose cost is at most $\alpha$ times the optimal cost for every instance of the problem. This guarantee $\alpha$ is also referred to as the approximation factor of the algorithm. Currently, there are no constant factor approximation algorithms for the ATSP even when the costs satisfy the triangle inequality. The approximation factors of the

1. Graduate Student, Mechanical Engineering, Texas A & M University, College Station, TX 77843.
2. Assistant Professor, Mechanical Engineering, Texas A & M University, College Station, TX 77843. srathinam@tamu.edu.

existing algorithms for the ATSP either depend on the number of targets [7],[8],[9] or the input data[7]. For example, the well known covering algorithm for the ATSP in [7] has an approximation factor of $\log(n)$ where $n$ is the number of targets. There are also data dependent algorithms [7] with the approximation factors that depend on $\max_{i,j} \frac{c_{ij}}{c_{ji}}$ where $c_{ij}$ denotes the cost of traveling from vertex $i$ to vertex $j$.

When the travel costs are *symmetric* and satisfy the triangle inequality, authors in [10] provide an approximation algorithm for the FCURP. They assume that the minimum fuel required to travel from any target to its nearest depot is at most equal to $\frac{La}{2}$ units where $a$ is a constant in the interval $[0,1]$ and $L$ is the fuel capacity of the vehicle. This is a reasonable assumption, as in any case, one cannot have a feasible tour if there is a target that cannot be visited from any of the depots. Using these assumptions, Khuller et al. [10] present a $\frac{3(1+a)}{2(1-a)}$-approximation algorithm for the problem. In this article, we generalize this result for the asymmetric case.

FCURP is related to a more general search problem with uncertainties [11] where the fuel constraints are posed as a restriction on the time spent by the vehicle between any two successive depots on its path. The authors in [11] discretize time and space, and develop heuristics based on the shortest path algorithms. There are also variants of the vehicle routing problem that are closely related to the FCURP. For example, in [12], [13], the authors address a symmetric version of the arc routing problem where there is a single depot and a set of intermediate facilities, and the vehicle has to cover a subset of edges along which customers are present. The vehicle is required to collect goods from the customers as it traverses the given set of edges and unload the goods at the intermediate facilities. The goal of this problem is to find a tour of minimum length that starts and ends at the depot such that the vehicle visits the given subset of edges and the total amount of goods carried by the vehicle never exceeds the capacity of the vehicle at any location along the tour. One of the key differences between the arc routing problem and the FCURP is that there is no requirement that any subset of edges must be visited in the FCURP. There are also similar problems [14], [15] addressed in the literature where each customer is located at a distinct vertex (instead of being present along the edges) and the vehicle is required to collect goods from the customers and deliver them at the intermediate facilities. FCURP is also different from the single depot vehicle routing problems addressed in [16], [17], [18] where there are additional length, travel-time or capacity constraints.

In the context of the above results, the following are the contributions of this article for the FCURP:
1. An algorithm with an approximation factor of $\frac{(1+a+2a\beta)\log(|T|)}{(1-a)}$ where $T$ represents the set of targets, and $a$ and $\beta$ are data dependent constants (presented in section III).
2. Fast construction and improvement heuristics to improve upon the solutions found by the approximation algorithm (presented in section IV).
3. A mixed-integer linear program to find an optimal solution for the FCURP (presented in section V). This optimal solution will then be used to corroborate the quality of solutions produced by the approximation algorithm and the heuristics.
4. Computational results to compare the performance of all the algorithms with respect to the quality of the solutions produced by the algorithms and their respective computation times (presented in section VI).

## II. PROBLEM STATEMENT

Let $T$ denote the set of targets and $D$ represent the set of depots. Let $s \in D$ be the depot where the UAV is initially located. The FCURP is formulated on the complete directed graph $G = (V, E)$ with $V = T \cup D$. Let $f_{ij}$ represent the amount of fuel required by the vehicle to travel from vertex $i \in V$ to vertex $j \in V$. It is assumed that the fuel costs satisfy the triangle inequality *i.e.*, for all distinct $i, j, k \in V$, $f_{ij} + f_{jk} \geq f_{ik}$.

Let $L$ denote the maximum fuel capacity of the vehicle. For any given target $i \in T$, we will assume that there are depots $d_1$ and $d_2$ such that $f_{d_1 i} + f_{i d_2} \leq aL$ where $a$ is a fixed constant in the interval $[0, 1]$. This is a reasonable assumption, as in any case, target $i$ cannot be visited by the vehicle if there are no depots $d_1$ and $d_2$ such that $f_{d_1 i} + f_{i d_2} > L$. We will also assume that it is always possible to travel from one depot to any another depot (either directly or by passing through some intermediate depots) without violating the fuel constraints. Given two distinct depots $d_1$ and $d_2$, let $l'_{d_1, d_2}$ denote the minimum fuel required to travel from $d_1$ to $d_2$. Then, let $\beta$ be a constant such that $l'_{d_2, d_1} \leq \beta l'_{d_1, d_2}$ for all distinct $d_1, d_2 \in D$.

A path for the vehicle is denoted by a sequence of vertices $(v_1, v_2, \cdots, v_k)$ visited by the vehicle where $v_i \in V$ for $i = 1, \cdots, k$. A tour for the vehicle is defined as a path that starts and terminates at the same vertex. The travel cost associated with any collection of edges present in the tour is defined as the sum of the fuel required to travel all the edges in the collection. Without loss of generality, we will assume that there is a target exactly at the location of the initial depot; therefore, a tour visiting all the targets can be transformed to a tour visiting all the targets and the initial depot and vice versa.

The objective of the problem is to find a tour such that
- the tour starts and terminates at the initial depot,
- the UAV visits each target at least once,
- the fuel required to travel any part of the tour which starts at a depot, visits a subset of targets and ends at the next depot must be at most equal to $L$, and,
- the travel cost associated with the edges in the tour is a minimum.

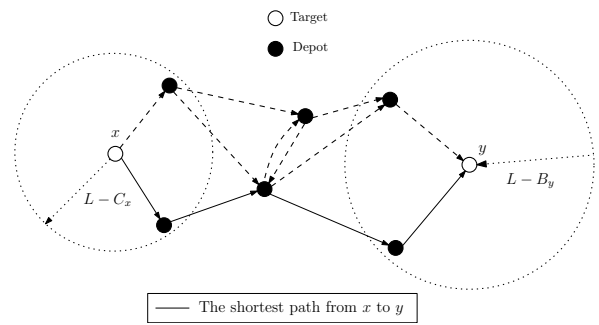

Fig. 2. The first step of the approximation algorithm: The solid edges represent the shortest path $PATH_{xy}$ from target $x$ to target $y$, and the cost of traveling this path is denoted by $l_{xy}$.

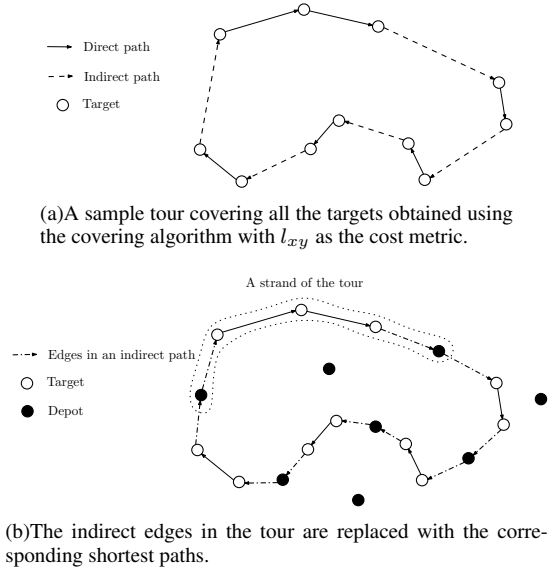

(a) A sample tour covering all the targets obtained using the covering algorithm with $l_{xy}$ as the cost metric.

(b) The indirect edges in the tour are replaced with the corresponding shortest paths.

Fig. 3. An illustration of the second step of the approximation algorithm.

## III. Approximation Algorithm

We refer to this approximation algorithm as $Approx$. There are three main steps in $Approx$. The **first step** of $Approx$ aims to find a path for the vehicle to travel from any target $x \in T$ to any other target $y \in T$ such that the path can be a part of a feasible tour for the FCURP, the path satisfies all the refueling constraints and the travel cost associated with the path is a minimum. Note that the maximum amount of fuel available for the vehicle when it reaches target $x$ in any tour is $L - \min_d f_{dx}$. Also, in any feasible tour, there must be at least $\min_d f_{yd}$ units of fuel left when the vehicle reaches target $y$ so that the vehicle can continue to visit other vertices along its tour. Define $C_x := \min_d f_{dx}$ and $B_x := \min_d f_{xd}$ for any $x \in T$. The first step of the $Approx$ essentially finds a feasible path of least cost (also referred as the shortest path) such that the vehicle starts at target $x$ with at most $L - C_x$ units of fuel and ends at target $y$ with at least $B_y$ units of fuel. If there is enough fuel available for the vehicle to travel from $x$ to $y$ (or, if $L - C_x - B_y \geq f_{xy}$), the vehicle can directly reach $y$ from $x$ while respecting the fuel constraints. In this case, we say that the vehicle can *directly* travel from $x$ to $y$ and the shortest path (also referred to as the *direct* path) is denoted by $PATH(x,y) := (x,y)$. The cost of traveling this shortest path is just $f_{xy}$.

If the vehicle *cannot directly* travel from $x$ to $y$ (if $L - C_x - B_y < f_{xy}$), the vehicle must visit some of the depots on the way before reaching target $y$. In this case, we find a shortest path using an auxiliary directed graph, $(V', E')$, defined on all the depots and the targets $x, y$, i.e., $V' = D \cup \{x, y\}$ (illustrated in figure 2). An edge is present in this directed graph only if traveling the edge can satisfy the fuel constraint. For example, as the vehicle has at most $L - C_x$ units of fuel to start with, the vehicle can reach a depot $d$ from $x$ only if $f_{xd} \leq L - C_x$. Therefore, $E'$ contains an edge $(x, d)$ if the constraint $f_{xd} \leq L - C_x$ is satisfied. Similarly, the vehicle can travel from a depot $d$ to target $y$ only if there are at least $B_y$ units of fuel remaining after the vehicle reaches $y$. Therefore, $E'$ contains an edge $(d, y)$ if the constraint $f_{dy} \leq L - B_y$ is satisfied. In summary, the following are the edges present in $E'$:

$$E' := \begin{cases} \{(x,d) : \forall d \in D, f_{xd} \leq L - C_x\}, \\ \bigcup \{(d_1, d_2) : \forall d_1, d_2 \in D, f_{d_1 d_2} \leq L\}, \\ \bigcup \{(d, y) : \forall d \in D, f_{dy} \leq L - B_y\}. \end{cases} \quad (1)$$

Any path starting at $x$ and ending at $y$ in this auxiliary graph will require the vehicle to carry at most $L - C_x$ units of fuel at target $x$, satisfy all the fuel constraints and reach target $y$ with at least $B_y$ units of fuel left. Also, we let the cost of traveling any edge $(i, j) \in E'$ to be equal to $f_{ij}$ (as defined in section II). Now, we use Dijkstra's algorithm [19] to find a shortest path to travel from $x$ to $y$. This shortest path (also referred to as the *indirect* path using intermediate depots) can be represented as $PATH(x, y) := (x, d_1, d_2, \cdots, y)$.

In the **second step** (illustrated in figure 3) of $Approx$, we use the shortest path computed between any two targets to find a tour for the vehicle. To do this, let $l_{xy}$ denote the cost of the shortest path $PATH(x, y)$ that starts at $x$ and ends at $y$. The following covering algorithm [7] is used to obtain a tour which visits each of the targets at least once. Suppose $G'_o$ represent the collection of edges chosen by the covering algorithm. Initially, $G'_o$ is an empty set.

- Let $T' := T$. Find a minimum cost cycle cover, $C$, for the graph $(T', E'_T)$ with $E'_T := \{(x, y) : x, y \in T'\}$ and $l_{xy}$ as the cost metric. A cycle cover for a graph is a collection of edges such that the indegree and the outdegree of each vertex in the graph is exactly equal to one. A minimum cost cycle cover is a cycle cover such that the sum of the cost of the edges in the cycle cover is a minimum. This step can be solved in at most $O(|T'|^3)$ steps using the Hungarian algorithm [7]. Add all the edges found in $C$ to $G'_o$.
- If the cycle cover consists of at least two cycles, select exactly one vertex from each cycle and return to step 1 with $T'$ containing only the selected vertices. If the cycle cover $C$ consists of exactly one cycle go to the next step.
- The collection of edges in $G'_o$ represents a connected Eulerian graph spanning all the targets where the indegree and the outdegree of each target is the same. Given an Eulerian graph, using Euler's theorem, one can always find a tour such that each edge in $G'_o$ is visited exactly once. This tour is the output of the covering algorithm.

If there is any edge $(x, y)$ in this tour such that the vehicle *cannot directly* travel from $x$ to $y$, $(x, y)$ is replaced with all the edges present in the shortest path, $PATH(x, y)$, from $x$ to $y$. After replacing all the relevant edges with the edges from the shortest paths, one obtains a Hamiltonian tour, $TOUR$, which visits each of the targets at least once and some of the intermediate depots for refueling. This tour may still be infeasible because there may be a sequence of vertices that starts at a depot and ends at the next depot on the tour which may not satisfy the fuel constraints. To correct this, we further augment this tour with more visits to the depots as explained in the next step of the algorithm.

In the **last step** of $Approx$ (illustrated in figure 2), the entire tour, $TOUR$, obtained from the second step is decomposed into a series of strands. A strand is a sequence of adjacent vertices in the tour that starts at a depot, visits a set of targets and ends at a depot. $TOUR$ must be infeasible if the total fuel required to travel any one of these strands is greater than the fuel capacity of the vehicle ($L$). Hence, in this step, all the infeasible strands are identified, and a greedy algorithm is applied to each infeasible strand to transform it to a feasible strand (refer to figure 4). We present some definitions before we outline the greedy algorithm. A depot, $m_x$, is referred as a *nearest starting* depot for $x$ if $f_{m_x x} = \min_d f_{dx}$. Similarly, a depot $n_x$ is referred as a *nearest terminal depot* for $x$ if $f_{x n_x} = \min_d f_{xd}$. As in the second step of the algorithm, given any two depots $d_s, d_f \in D$, one can find a path of least cost that starts from $d_s$, visits some intermediate depots (if necessary) and ends at $d_f$ while satisfying all the fuel constraints [1]. Let the sequence of all the depots in this path be denoted by $PATH(d_s, d_f) := (d_s, \bar{d}_1, \bar{d}_2, \cdots, \bar{d}_k, d_f)$ where $\bar{d}_1, \bar{d}_2, \cdots, \bar{d}_k \in D$ are the intermediate depots visited by the vehicle.

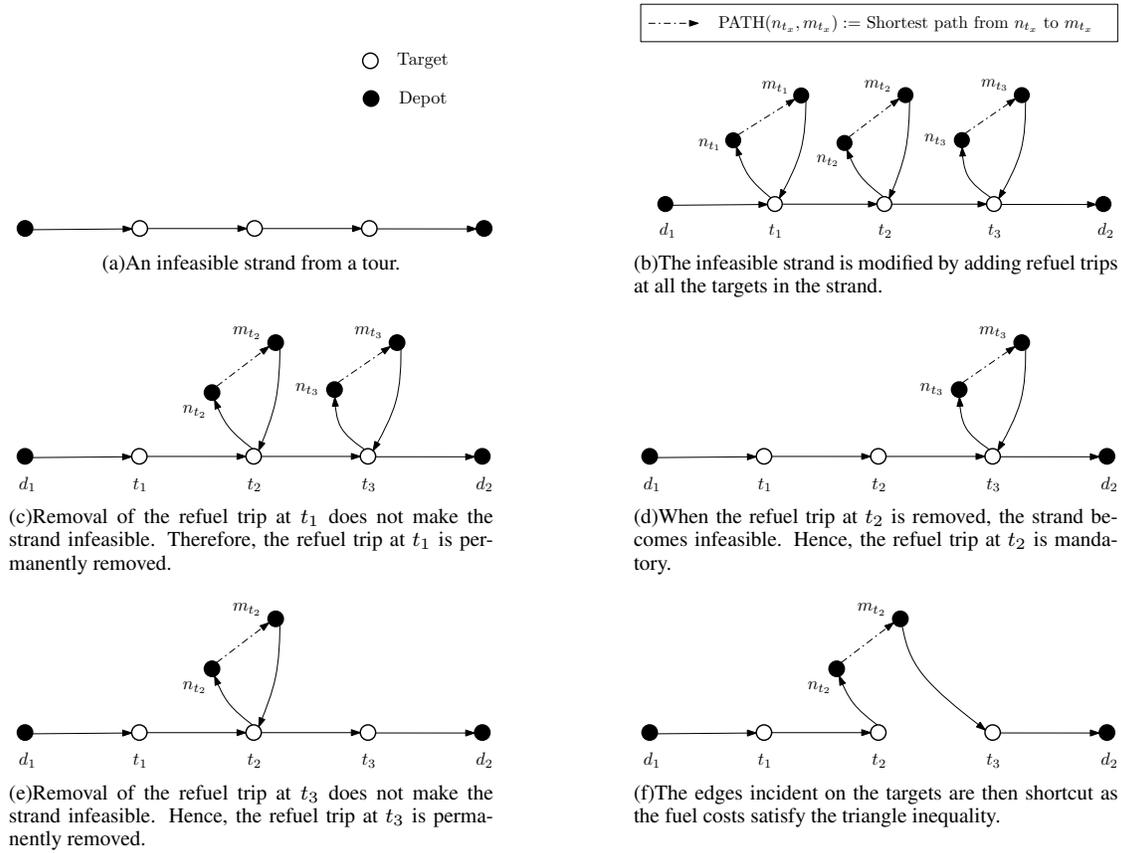

Fig. 4. The greedy procedure to convert an infeasible strand to a feasible strand.

The greedy algorithm works as follows (refer to figure 4): Consider an infeasible strand represented as $(d_1, t_1, \cdots, t_k, d_2)$ where $d_1$ and $d_2$ are the two depots of the strand and $t_1, \cdots, t_k$ are the targets. For each target $t$ in this infeasible strand, we add a refueling trip such that

- The vehicle visits a nearest terminal depot $n_t$ after leaving $t$.
- The vehicle uses the sequence of depots specified in $PATH(n_t, m_t)$ to travel from $n_t$ to $m_t$ where $m_t$ is the nearest

[1] Apply Dijkstra's algorithm on the graph $(D, E_d)$ where $E := \{(i,j) : i, j \in D, f_{ij} \leq L\}$ and the cost of traveling from vertex $i \in D$ to vertex $j \in D$ is $c_{ij}$.

starting depot for $t$, and finally returns to $t$ after refueling.

After adding all the refueling trips, the modified strand can be denoted as $(d_1, t_1, PATH(n_{t_1}, m_{t_1}), t_1, t_2, PATH(n_{t_2}, m_{t_2}), t_2, \ldots, PATH(n_{t_k}, m_{t_k}), t_k, d_2)$. Now, each of the refueling trips is chosen sequentially in the order they are added and is shortcut if the strand that results after removing the refueling trip still satisfies the fuel constraint (refer to figure 4).

### A. Analysis of the Approximation Algorithm

*Lemma III.1:* $Approx$ always produces a feasible solution for the FCURP.

*Proof:* Consider the greedy procedure presented in the last step of the $Approx$ which attempts to convert an infeasible strand $(d_1, t_1, t_2, \cdots, t_k, d_2)$ into a feasible path for the vehicle. The edges $(d_1, t_1)$ and $(t_k, d_2)$ in this strand belong to indirect paths while the remaining edges belong to direct paths. The vehicle can always travel from $d_1$ to $t_1$ and still have enough fuel at $t_1$ to reach its nearest terminal depot as edge $(d_1, t_1)$ was added according to the fuel constraints in (1). Therefore, once the vehicle reaches $t_1$, due to our assumptions on the fuel costs, there always exists a refueling trip such that the vehicle starts at $t_1$, visits the depots $n_{t_1}, m_{t_1}$ before returning to $t_1$ with the maximum amount of fuel possible at $t_1$. As a result, the vehicle must be able to reach $t_2$ with sufficient amount of fuel remaining to reach the nearest terminal depot $n_{t_2}$. Again, there exists a refueling trip such that at the end of this trip, the vehicle can return to $t_2$ with the maximum amount of fuel possible at $t_2$. The above arguments can be repeatedly used for each target in the infeasible strand to show that the vehicle must be able to reach $d_2$ using the modified strand while satisfying the fuel constraints. Therefore, the greedy procedure can always convert any infeasible strand into a feasible path and hence, the $Approx$ finds a feasible solution to the FCURP. ∎

The cost of the final solution (say $TOUR_f$) obtained by $Approx$ is upper bounded by the sum of the cost of $TOUR$ and the cost of all the refueling trips. So, in order to bound the cost of $TOUR_f$, we need to bound the cost of $TOUR$, the number of refueling trips and the cost of each refueling trip in terms of the optimal cost of the FCURP. In the following lemma, we first bound the cost of $TOUR$.

*Lemma III.2:* Let $cost(TOUR)$ denote the total fuel required to travel all the edges in $TOUR$. Then, $cost(TOUR)$ is at most equal to $\log(|T|) \times C_{opt}$ where $C_{opt}$ is the optimal cost of the FCURP.

*Proof:* The cost of $TOUR$ is equal to the sum of the cost of all the cycle covers spanning all the targets with $l_{xy}$ as the cost metric. Now, consider any minimum cost cycle cover $C$ spanning the targets $t_1, t_2, \cdots, t_m$. Without loss of generality, we also let $(t_1, t_2, \cdots, t_m)$ denote the sequence in which the targets in $C$ are visited in an optimal solution to the FCURP. The minimum cost, $l_{t_i, t_{i+1}}$, of traveling from target $t_i$ to $t_{i+1}$ (computed in the first and the second step of $Approx$) must be at most equal to the cost of traveling from $t_i$ to $t_{i+1}$ in the optimal solution of the FCURP. Therefore, the minimum cost of a TSP tour visiting any subset of targets using $l_{xy}$ as the metric must be at most equal to $C_{opt}$. Since the problem of computing a minimum cost cycle cover is a relaxation to the TSP, it follows that the cost of any optimal cycle cover computed in the second step of $Approx$ must be at most equal to $C_{opt}$. The number of iterations in the covering algorithm is at most $\log(|T|)$ as the number of selected targets in any two successive iterations of the covering algorithm reduces by half. Hence, the cost of $TOUR$ which is the same as the total cost of all the cycle covers is at most equal to $\log(|T|) \times C_{opt}$. ∎

In the following lemma, we bound the number of refueling trips needed to make $TOUR$ feasible.

*Lemma III.3:* The number of refueling trips needed by the vehicle is upper bounded by $\frac{2cost(TOUR)}{(1-a)L}$.

*Proof:* Let $I = (d_1, t_1, t_2, \cdots, d_2)$ represent an infeasible strand in $TOUR$ that requires additional refueling trips and let $cost(I)$ denote the total fuel required to travel the edges connecting any two adjacent vertices in $I$. Given any two vertices $u, v \in I$ and the segment $I_{uv}$ of $I$ starting at $u$ and ending at $v$, let $cost(u, v)$ denote the total fuel required to travel the edges connecting any two adjacent vertices in $I_{uv}$. Let the greedy procedure add refueling trips at targets $v_1, v_2, \cdots, v_k$ to make $I$ feasible. Then, $cost(d_1, v_2)$ must be greater than $L - B_{v_2}$ (recall that for any target $x$, $C_x := \min_d f_{dx}$ and $B_x := \min_d f_{xd}$); if this is not the case, the refueling trip at target $v_1$ is unnecessary and can be removed. Similarly, $cost(v_1, v_3)$ must be greater than $L - C_{v_1} - B_{v_3}$, else, the refueling trip at $v_2$ is avoidable and can be removed. Repeating the above arguments for the pairs of vertices $(v_2, v_4), \cdots, (v_{k-2}, v_k)$ and $(v_{k-1}, d_2)$, we get, the following inequalities:

$$\begin{aligned} cost(d_1, v_2) &> L - B_{v_2}, \\ cost(v_1, v_3) &> L - C_{v_1} - B_{v_3}, \\ cost(v_2, v_4) &> L - C_{v_2} - B_{v_4}, \\ &\vdots \quad \vdots \quad \vdots \\ cost(v_{k-2}, v_k) &> L - C_{k-2} - B_{v_k}, \\ cost(v_{k-1}, d_2) &> L - C_{k-1}. \end{aligned}$$

Now, the number of refueling trips can be bounded in the following way:

$$\begin{aligned} 2cost(I) &\geq cost(d_1, v_2) + \sum_{i=1}^{k-2} cost(v_i, v_{i+2}) + cost(v_{k-1}, d_2) \\ &\geq kL - C_{v_1} - \sum_{x=v_2, \cdots, v_{k-1}} (B_x + C_x) - B_{v_k}. \quad (2) \end{aligned}$$

For any target $x$, as there are depots $\widehat{d}$ and $\overline{d}$ such that $f_{\widehat{d}x} + f_{x\overline{d}} \leq aL$, we have $C_x + B_x = \min_d f_{dx} + \min_d f_{xd} \leq f_{\widehat{d}x} + f_{x\overline{d}} \leq aL$. Using this bound for each target in (2), we get

$$\begin{aligned} 2cost(I) &\geq kL - aL - (k-2)aL - aL \quad (3) \\ &\geq k(1-a)L. \quad (4) \end{aligned}$$

As a result, the number of refueling trips for strand $I$ is upper bounded by $\frac{2cost(I)}{(1-a)L}$. Therefore, the total number of refueling trips for the infeasible strands is upper bounded by $\frac{2}{(1-a)L} \sum_I cost(I) \leq \frac{2}{(1-a)L} cost(TOUR)$. ∎

The following theorem provides an approximation factor for $Approx$ which depends on the size of the problem and the input data.

*Theorem III.1:* $Approx$ solves the FCURP with an approximation factor of $\frac{(1+a+a\beta)\log(|T|)}{1-a} L$ in $O(|D|^2|T|^2 + |T|^3 \log(|T|))$ steps.

*Proof:* The cost of the solution, $TOUR_f$, obtained by $Approx$ is upper bounded by the sum of the cost of $TOUR$ and the cost of all the refueling trips. Note that the cost of the refueling trip at any target $x$ must be equal to $f_{x\widehat{d}_1} + f_{\widehat{d}_1\widehat{d}_2} + f_{\widehat{d}_2 x}$ where the depots $\widehat{d}_1, \widehat{d}_2$ are such that $f_{x\widehat{d}_1} = \min_d f_{xd}$, $f_{\widehat{d}_2 x} = \min_d f_{dx}$. From the assumptions in section II, we get,

$$\begin{aligned} f_{x\widehat{d}_1} + f_{\widehat{d}_1\widehat{d}_2} + f_{\widehat{d}_2 x} &\leq f_{x\widehat{d}_1} + \beta f_{\widehat{d}_2\widehat{d}_1} + f_{\widehat{d}_2 x} \\ &\leq f_{x\widehat{d}_1} + \beta(f_{\widehat{d}_2 x} + f_{x\widehat{d}_1}) + f_{\widehat{d}_2 x} \\ &= (1+\beta)(f_{\widehat{d}_2 x} + f_{x\widehat{d}_1}) \\ &\leq (1+\beta)aL. \end{aligned}$$

Using lemma III.3, we can conclude that the total cost of all the refueling trips must be at most equal to $(1+\beta)aL \times \frac{2cost(TOUR)}{(1-a)L} = \frac{2(1+\beta)a}{(1-a)} cost(TOUR)$. Therefore, the total cost of $TOUR_f$ is upper bounded by $cost(TOUR) + \frac{2(1+\beta)a}{(1-a)} cost(TOUR) = \frac{(1+a+2\beta a)}{(1-a)} cost(TOUR)$. Using lemma III.2, we get, $cost(TOUR_f) \leq \frac{(1+a+2\beta a)}{(1-a)} \log(|T|) C_{opt}$. Also, the number of steps involved in the algorithm is dominated by the first and second step of $Approx$. For any given

pair of targets $x$ and $y$, the Dijkstra's algorithm requires at most $O(|D|^2)$ steps to compute $l_{xy}$. As a result, the total number of steps required to implement the first step of $Approx$ is $O(|D|^2|T|^2)$. The second step of $Approx$ runs the Hungarian algorithm for at most $log(|T|)$ iterations. Hence, the number of steps required to implement the second step is $O(|T|^3 log(|T|))$. Therefore, the total number of steps involved in $Approx$ is $O(|D|^2|T|^2 + |T|^3 log(|T|))$. ∎

## IV. CONSTRUCTION AND IMPROVEMENT HEURISTICS

The construction heuristic we propose is exactly the same as $Approx$ except for its second step. Specifically, we replace the covering algorithm in the second step of $Approx$ with the Lin-Kernighan-Helgaun (LKH) heuristic [20]. We then use the solution obtained using the construction heuristic as an initial feasible solution for the improvement heuristics. The improvement heuristics relies on a combination of a $k-$opt heuristic and a depot exchange heuristic to improve the quality of the tour obtained by the construction heuristic. A $k-$opt heuristic is a local search method which iteratively attempts to improve the quality of a solution until some termination criteria are met. The depot exchange heuristic aims to replace some depots in the tour with refueling depots not present in the tour in order to obtain better feasible solutions. A flow chart of the overall procedure is presented in figure 5. In the following subsections, we explain the $k-$opt and the depot exchange heuristic in detail.

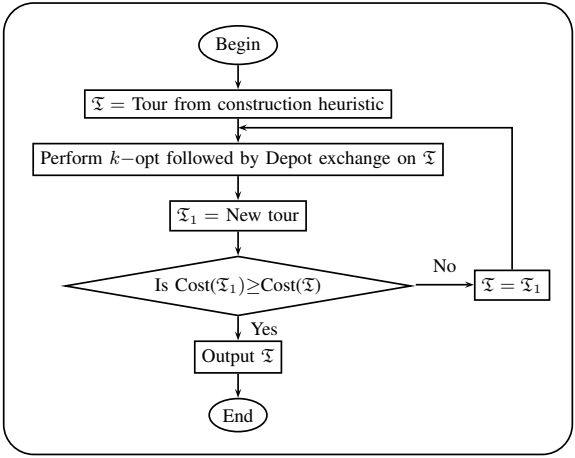

Fig. 5. Overall procedure in the improvement heuristic.

### A. $k-$opt

We will first give some basic definitions involved in a $k-$opt heuristic, and then see how it is applicable to the FCURP. A tour $S_2$ is defined to be in the $k-$exchange neighborhood of the tour $S_1$ if $S_2$ can be obtained from $S_1$ by replacing $k$ edges in $S_1$ with $k$ new edges. A tour $S_2$ is said to be obtained from a feasible tour $S_1$ by an improving $k'-$exchange if $S_2$ is in the $k'-$exchange neighborhood of $S_1$, is feasible and has a travel cost lower than $S_1$. The $k-$opt heuristic starts with a feasible tour and iteratively improves on this tour making successive improving $k'-$exchanges for any $2 \leq k' \leq k$ until no such exchanges can be made.

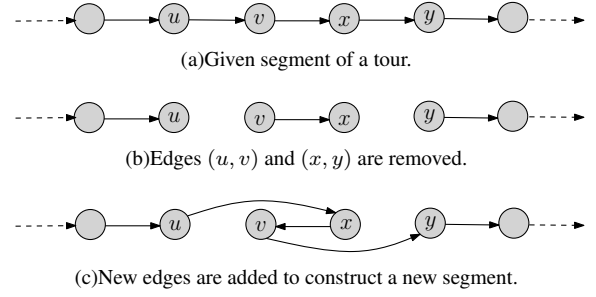

(a) Given segment of a tour.

(b) Edges $(u,v)$ and $(x,y)$ are removed.

(c) New edges are added to construct a new segment.

Fig. 6. Possible 2-exchange move.

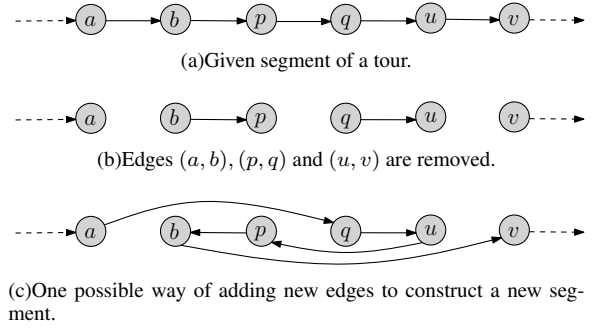

(a) Given segment of a tour.

(b) Edges $(a,b)$, $(p,q)$ and $(u,v)$ are removed.

(c) One possible way of adding new edges to construct a new segment.

Fig. 7. Possible 3-exchange move.

---

**Algorithm 1** : Pseudo code for the $k-$opt algorithm

*Notations:* Let $cost(\mathfrak{T})$ denote the sum of the cost of traveling all the edges in the tour $\mathfrak{T}$. Let $n$ denote the *search span* of a segment.

1: $\mathfrak{T}^* \leftarrow$ Initial feasible tour.
2: $\mathfrak{T} \leftarrow \mathfrak{T}^*$.
3: **loop**
4:   $N_d \leftarrow$ Number of visits to the depots in $\mathfrak{T}$.
5:   **for** $i = 1, \cdots, N_d$ **do**
6:     $\mathfrak{S}(i,n) \leftarrow$ segment of $\mathfrak{T}$ centered at the $i^{th}$ depot visited in $\mathfrak{T}$.
7:     Find a tour $\mathfrak{R}$ such that for $2 \leq k' \leq k$,
8:       $\mathfrak{R}$ is obtained by replacing $k'$ edges in the segment $\mathfrak{S}(i,n)$ with $k'$ new edges;
9:       $\mathfrak{R}$ is the best improving $k'-$exchange of $\mathfrak{T}$.
10:     **If** $cost(\mathfrak{R}) < cost(\mathfrak{T})$, $\mathfrak{T} \leftarrow \mathfrak{R}$.
11:   **end for**
12:   **if** $cost(\mathfrak{T}^*) \leq cost(\mathfrak{T})$ **then**
13:     break;
14:   **else**
15:     $\mathfrak{T}^* \leftarrow \mathfrak{T}$.
16:   **end if**
17: **end loop**
18: Output $\mathfrak{T}^*$ as the solution.

A critical part of developing a $k-$opt heuristic deals with choosing an appropriate $k'-$exchange neighborhood for a tour. One way to choose this is to consider all possible subsets of $k'$ edges in the tour and try an improving $k'-$exchange. Initial implementations showed us that substantial improvements in the quality of the tour were obtained when the $k'-$exchanges where performed around the refueling depots in the tour. In view of this observation, we define a segment of span $n$ as a sequence of $2n+1$ adjacent vertices of the tour centered around a depot. A segment can be denoted by $(s_1, \ldots, s_n, d, s_{n+1}, \ldots, s_{2n})$, where $d$ is the depot around which the segment is centered. Following the definition of a segment, one can infer that the number of possible segments in a feasible tour is equal to the number of visits by the UAV to all the depots.

The $k'-$exchange neighborhood in each iteration is restricted to one of the segments of the given tour. Given a segment, $k'$ edges are deleted from the segment, and subsequently $k'$ new edges are added to form a new segment as shown in figures 6 and 7. The updated tour is then checked for feasibility to ensure that the UAV never runs out of fuel. The pseudo code for the $k-$opt heuristic is shown in algorithm 1. An illustration for $2-$opt and $3-$opt is shown in Figures 6 and 7.

### B. Depot Exchange Heuristic

Given a tour, we consider the depots in the order in which they are visited by the UAV and substitute each of them with a (possibly) new refueling depot in order to obtain a better feasible solution. For a given depot $d$ in the tour, suppose $v_1$ and $v_2$ are the vertices that are visited immediately before and after visiting $d$ in the tour. The heuristic replaces $d$ with a new depot $d_r := \mathrm{argmin}_{u \in D}\ c_{v_1 u} + c_{u v_2}$ if the new tour is feasible and reduces the total cost. The new tour then acts as the current feasible solution and the above procedure is repeated for each depot until no further improvements can be done.

## V. MIXED INTEGER PROGRAMMING FORMULATION

Let $x_{ij}$ denote an integer decision variable which determines the number of directed edges from vertex $i$ to $j$ in the network; that is, $x_{ij}$ is equal to $q$ if and only if the vehicle travels $q$ times from vertex $i$ to vertex $j$. As the costs satisfy the triangle inequality, without loss of generality, we can assume that there is an optimal solution such that each target is visited exactly once by the vehicle. Therefore, we restrict $x_{ij} \in \{0,1\}$ if either vertex $i$ or vertex $j$ is a target.

The collection of edges chosen by the formulation must reflect the fact that there must be a path from the depot to every target. We use flow constraints [21] to formulate this connectivity constraint. In these flow constraints, the vehicle collects $|T|$ units of a commodity at the depot and delivers one unit of commodity at each target as it travels along its path. Enforcing that these commodities can be routed through the chosen edges ensures there is a path from the depot to every target. Suppose $p_{ij}$ denotes the amount of commodity flowing from vertex $i$ to vertex $j$. Also, let $r_i$ represent the fuel left in the vehicle when the $i^{th}$ target is visited. The FCURP can be formulated as a mixed integer linear program as follows:

$$\min \sum_{(i,j) \in E} c_{ij}\, x_{ij}$$

subject to *Degree constraints:*

$$\sum_{i \in V \setminus \{k\}} x_{ik} = \sum_{i \in V \setminus \{k\}} x_{ki} \quad \forall k \in V, \tag{5}$$

$$\sum_{i \in V \setminus \{k\}} x_{ik} = 1 \quad \forall k \in T, \tag{6}$$

*Capacity and flow constraints:*

$$\sum_{i \in V \setminus \{s\}} (p_{si} - p_{is}) = |T|, \tag{7}$$

$$\sum_{j \in V \setminus \{i\}} (p_{ji} - p_{ij}) = 1 \quad \forall i \in T, \tag{8}$$

$$\sum_{j \in V \setminus \{i\}} (p_{ji} - p_{ij}) = 0 \quad \forall i \in D \setminus \{s\}, \tag{9}$$

$$0 \leq p_{ij} \leq |T| x_{ij} \quad \forall i, j \in V, \tag{10}$$

*Fuel constraints:*

$$r_j - r_i + f_{ij} \leq M(1 - x_{ij}) \quad \forall i, j \in T, \tag{11}$$
$$r_j - r_i + f_{ij} \geq -M(1 - x_{ij}) \quad \forall i, j \in T, \tag{12}$$
$$r_j - L + f_{ij} \geq -M(1 - x_{ij}) \quad \forall i \in D \text{ and } j \in T, \tag{13}$$
$$r_j - L + f_{ij} \leq M(1 - x_{ij}) \quad \forall i \in D \text{ and } j \in T, \tag{14}$$
$$r_i - f_{ij} \geq -M(1 - x_{ij}) \quad \forall i \in T \text{ and } j \in D, \tag{15}$$

$$0 \leq r_i \leq L \quad \forall i \in T,$$
$$x_{ij} \in \{0,1\} \quad \forall i, j \in V, \text{ either } i \text{ or } j \text{ is a target},$$
$$x_{ij} \in \{0,1,2,\cdots,|T|\} \quad \forall i, j \in D. \tag{16}$$

Equation (5) states that the in-degree and out-degree of each vertex must be the same, and equation (6) ensures that each target is visited once by the vehicle. Note that these equations allow for the vehicle to visit a depot any number of times for refueling. The constraints in (7)-(10) ensure that there are $|T|$ units of commodity shipped from the depot and the vehicle delivers exactly one of commodity at each target. In equations, (11)-(15), $M$ denotes a large constant and can be chosen to be equal to $L + \max_{i,j \in V} f_{i,j}$. If the UAV is traveling from target $i$ to target $j$, equations (11) and (12) ensure that the fuel left in the vehicle after reaching target $j$ is $r_j = r_i - f_{ij}$. If the UAV is traveling from depot $i$ to target $j$, equations (13), (14) ensure that the fuel left in the vehicle after reaching target $j$ is $r_j = L - f_{ij}$. If the UAV is directly traveling from any target to a depot, constraint (15) states that the fuel remaining at the target must be at least equal to the amount required to reach the depot.

## VI. COMPUTATIONAL RESULTS

We considered problems of size ranging from 10 targets to 25 targets with increments in steps of 5. For each problem size,

50 instances were generated and all the targets were chosen randomly from a square area of $5000 \times 5000$ units. In addition, all the instances of the problem have 5 depots chosen at fixed locations in the square area. All the simulations were run on a Dell Precision T5500 workstation (Intel Xeon E5630 processor @ 2.53GHz, 12GB RAM).

The simulations were performed for a fixed wing vehicle with minimum turning radius constraints. A vehicle traveling at a constant speed with a bound on its turning radius is referred to as the Dubins' vehicle [6]. In the simulations, the minimum turning radius of the vehicle is chosen to be 100 units and the angle of approach for each target is selected uniformly in the interval $[0, 2\pi]$ radians. Given the approach angles at the targets, the minimum distance required to travel between any two targets subject to the turning radius constraints of the vehicle was solved by Dubins in [6]. For the simulations, the maximum fuel capacity $L$ was $4500$ units and we assumed that the fuel spent is directly proportional to the distance traveled by the vehicle.

The formulation presented in section V was solved to optimality using IBM ILOG CPLEX optimization software [22]. The average time required to find an optimal solution in CPLEX was nearly 2 hours for problem instances with 25 targets and 5 depots. On the other hand, the average time required to find a feasible solution using the approximation algorithm and the heuristics was less than 2 seconds for each tested instance. The quality of a solution produced by applying an algorithm on an instance $I$ is defined as $100 \times \frac{C_I^{algorithm} - C_I^{optimal}}{C_I^{optimal}}$ where $C_I^{algorithm}$ is the cost of the solution found by the algorithm and $C_I^{optimal}$ is the cost of the optimal solution for an instance $I$. The approximation algorithm and the heuristics were coded using Python 2.7.2 [23]. We used a search span of 4 for the improvement heuristics as it gave a good trade off between the solution quality and the computation time available.

The average quality of the solutions produced by the approximation algorithm and the heuristics for the instances is shown in the figure 9. From the figure, it is clear that the average quality of the solutions produced by the improvement heuristic is much superior compared to the average quality of the solutions found by the construction heuristic or the approximation algorithm. The depot-exchanges played a substantial part in improving the quality of the solutions found by the improvement heuristics; in particular, on an average, the depot exchange improved the solution quality by 0.14%, 0.66%, 0.78% and 1.10% for problems with 10, 15, 20 and 25 targets respectively. The feasible solution produced by the improvement heuristic was also used as an initial feasible solution for the formulation in CPLEX. The formulation was then solved in CPLEX with a time bound of 10 seconds. Using the feasible solution produced by the heuristic as a starting point, CPLEX was able to further improve the quality of the solutions as shown in figure 9. Specifically, for instances with 25 targets and 5 depots, CPLEX was further able to improve the average solution quality of the instances to 1.39%. These computational results show that the proposed algorithms can be effectively used in conjunction with standard optimization software like CPLEX in order to obtain high quality solutions for the FCURP. Considering that the FCURP is a difficult problem to solve, these results indicate that the approach proposed in this article is promising. Figures 8 show the paths found by the proposed algorithms for a Dubins' instance.

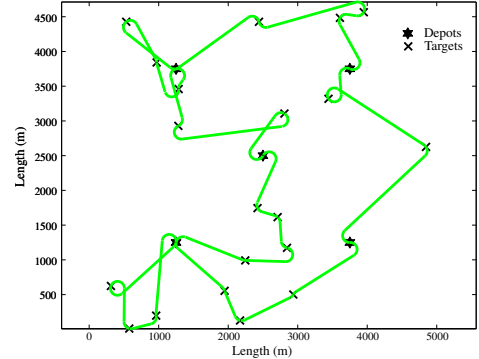

(a) Optimal solution

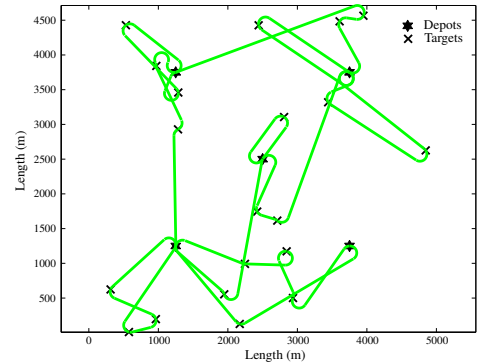

(b) Approximation algorithm

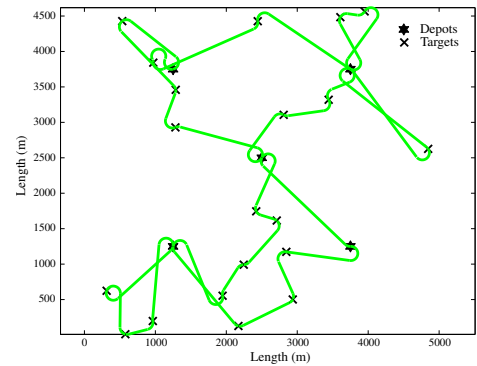

(c) Solution found by the improvement heuristic

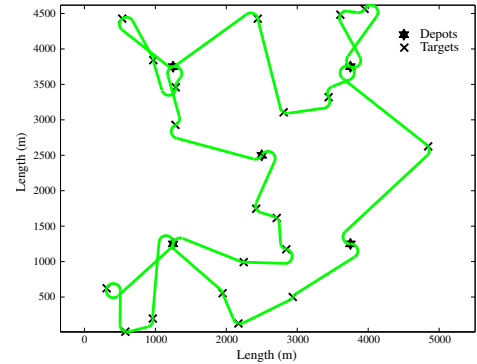

(d) Solution found by using CPLEX after 10 seconds with the solution in (b) as an initial feasible solution

Fig. 8. The paths found by the algorithms for a Dubins' instance with 25 nodes.

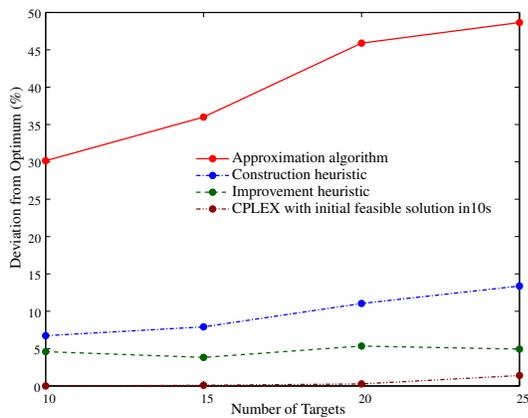

Fig. 9. Average quality of solutions produced by the proposed algorithms.

## VII. Conclusions

An approximation algorithm and fast heuristics were developed to solve a generalization of the single vehicle routing problem with fuel constraints. A mixed-integer, linear programming formulation was also proposed to find optimal solutions to the problem. Future work can be directed towards developing branch and cut methods, and can address problems with multiple, heterogeneous vehicles.